\documentclass[longbibliography,12pt]{iopart}
\usepackage{multirow}
\usepackage{graphicx}
\usepackage{epstopdf}
\usepackage{bm}
\usepackage{cite}

\begin{document}
\title{Superconducting phase above room temperature in lutetium-beryllium hydrides at high pressures }

\author{Bin Li$^{1}$, Yeqian Yang$^2$, Yuxiang Fan$^{1}$, Cong Zhu$^2$, Shengli Liu$^{1}$, Zhixiang Shi$^3$}
\address{$^1$ School of Science, Nanjing University of Posts and Telecommunications, Nanjing 210023, China}
\address{$^2$ College of Electronic and Optical Engineering, Nanjing University of Posts and Telecommunications, Nanjing 210023, China}
\address{$^3$ School of Physics, Southeast University, Nanjing 211189, China}
\ead{libin@njupt.edu.cn (B Li)}

\begin{abstract}
  High-pressure structural search was performed on the hydrogen-rich compound LuBeH$_8$ at pressures up to 200 GPa. We found a $Fm\overline{3}m$ structure that exhibits stability and superconductivity above 100 GPa. Our phonon dispersion, electronic band structure, and superconductivity analyses in the 100-200 GPa pressure range reveal a strong electron-phonon coupling in LuBeH$_8$. While $T_{c}$ shows a decreasing trend as the pressure increases, with a superconducting critical temperature $T_c$ of 255 K at 200 GPa and a maximum $T_c$ of 355 k at 100 GPa. Our research has demonstrated the room-temperature superconductivity in $Fm\overline{3}m$-LuBeH$_8$, thus enriching the family of ternary hydrides. These findings provide valuable guidance for identifying new high-temperature superconducting hydrides.

\end{abstract}
\noindent{\bf Keywords:}
\noindent{\it Superconductivity, Hydride, High pressures, Room temperature\/}\\
\maketitle

\section{Introduction}

The search for room-temperature superconducting materials is widely regarded as the "holy grail" of condensed matter physics\cite{van2010discovery}. According to Bardeen-Cooper-Schrieffer (BCS) theory\cite{bardeen1957theory}, the superconducting critical temperature ($T_{c}$) is proportional to the Debye temperature, which is inversely proportional to its mass. Metallic hydrogen, being the lightest element, has a high Debye temperature and strong electron-phonon (\emph{e-ph}) coupling, which can lead to high-temperature superconductivity. However, due to the extremely high pressure required for metallic hydrogen synthesis, it is technically difficult to achieve. Therefore, researchers focused on metal hydrides instead. The metallization of hydrides can be achieved at lower pressures due to the "chemical pre-compression" effect of heavier elements\cite{ashcroft1968metallic}.

The search for high-temperature superconductors in hydrogen-rich compounds began after the theory of "chemical precompression" was proposed. Initially, researchers focused on natural binary hydrides, such as SiH$_4$, AlH$_3$, etc.\cite{eremets2008superconductivity,goncharenko2008pressure}. These studies were followed by the investigation of binary hydrides with new proportions, such as H$_3$S (maximum $T_{c}$ is 203 K\cite{drozdov2015conventional}), CaH$_6$ ($T_{c}$ is 210 K at 170 GPa\cite{zhang2022superconductivity}), YH$_6$ ($T_{c}$ is 220 K at 183 GPa\cite{kong2021superconductivity}) and LaH$_{10}$ ($T_{c}$ is 250-260 K at 170-200 GPa\cite{geballe2018synthesis,drozdov2019superconductivity}), and so on. After exploring almost all binary hydrides, research shifted to ternary hydrides. Ternary hydrides greatly expand the variety of phases by providing more element ratios and leading to the discovery of higher superconducting transition temperatures. For example, CaYH$_{12}$ ($T_{c}$ = 258 K\cite{liang2019potential} at 200 GPa), Li$_2$MgH$_{16}$ ($T_{c}$ = 473 K at 250 GPa\cite{sun2019route}), LaBH$_8$ ($T_{c}$ = 126-156 K at 50-55 GPa\cite{di2021bh,liang2021prediction}).

The most prominent high-pressure high-$T_{c}$ compounds are known as "superhydrides". Superhydrides have enveloping cage-shaped hydrogen-based lattices, wrapped in positively charged metal atoms. The most prominent metal atoms are rare earth elements including lanthanum, yttrium, and cerium. However, the lutetium hydride has not received good attention\cite{semenok2020distribution,peng2017hydrogen}. Lutetium and lanthanum have similar electronegativity, and can dissociate hydrogen molecules into atoms, the $f$-shell filled with lutetium superhydride is expected to carry high $T_{c}$. Up until a recent study, superconducting properties were observed around room temperature ($T_{c}$ = 294 K) in nitrogen-doped lutetium hydride at mild pressure of 10 Kbar\cite{dasenbrock2023evidence}. Unfortunately, despite the use of various methods, such as X-ray diffraction (XRD), elemental analysis, Raman spectroscopy, etc., its composition and structure have not be clarified. Furthermore, recent experimental and theoretical endeavors have reported the absence of near-ambient superconductivity in nitrogen-doped lutetium hydrides\cite{Ming2023,xing2023observation,sun2023effect,hilleke2023structure,ferreira2023search}, which is contrary to the original work by Dasenbrock\cite{dasenbrock2023evidence}. The existence of superconductivity in nitrogen-lutetium hydrides is a topic of debate.

 In this letter, we predict a new ternary room-temperature superconductor, LuBeH$_8$ (spacegroup: $Fm\overline{3}m$), by searching the stable structures of lutetium-beryllium-hydrogen system. We studied its phonon dispersion, electronic band structure, electron-phonon couplings, and superconducting critical temperatures. Its high symmetry facilitates its good superconductivity. Through our calculations, we found that LuBeH$_8$ remains stable at 100 GPa, and the superconducting critical temperature is as high as 355 K, which is already far beyond the room temperature.

\section{Methods}
We used in-house developed machine-learning-based crystal structure prediction package CRYSTREE\cite{wu2022phase,crystree2} to search the stable crystal structure of the Lu-Be-H (element ratio of 1:1:8) system at 100, 150 and 200 GPa. The results are verified by the graph theory assisted universal structure searcher MAGUS \cite{nwad128}. We then re-optimized the structures using the $ab$ $initio$ calculation of the Quantum Espresso(QE) package\cite{giannozzi2009quantum}, and calculate the phonon spectrum at different pressures using density functional perturbation theory (DFPT)\cite{baroni2001phonons}. The charge density and the wave function cutoff values are 600 Ry and 60 Ry, respectively. Electronic structure calculations were performed by using the method of full-potential linearization enhanced plane wave (FP-LAPW)\cite{blaha1990full} with Perdew Burke Ernzerhof (PBE) functional. VESTA\cite{momma2011vesta} was used to visualize the crystal structure. Fermi surfaces were visualized using Fermisurfer\cite{kawamura2019fermisurfer}. A 4$\times$4$\times$4 q grid and a 12$\times$12$\times$12 k point grid were selected to calculate electron phonon coupling and integration in the Brillouin zone. Dense 24$\times$24$\times$24 grids are used to evaluate precise electron-phonon interaction matrices. Finally, $T_{c}$ was calculated using the Allen-Dynes modified McMillan equation\cite{wang2012superconductive}.

\section{Results and discussion}

\begin{figure*}
\begin{center}
\includegraphics[width=16cm]{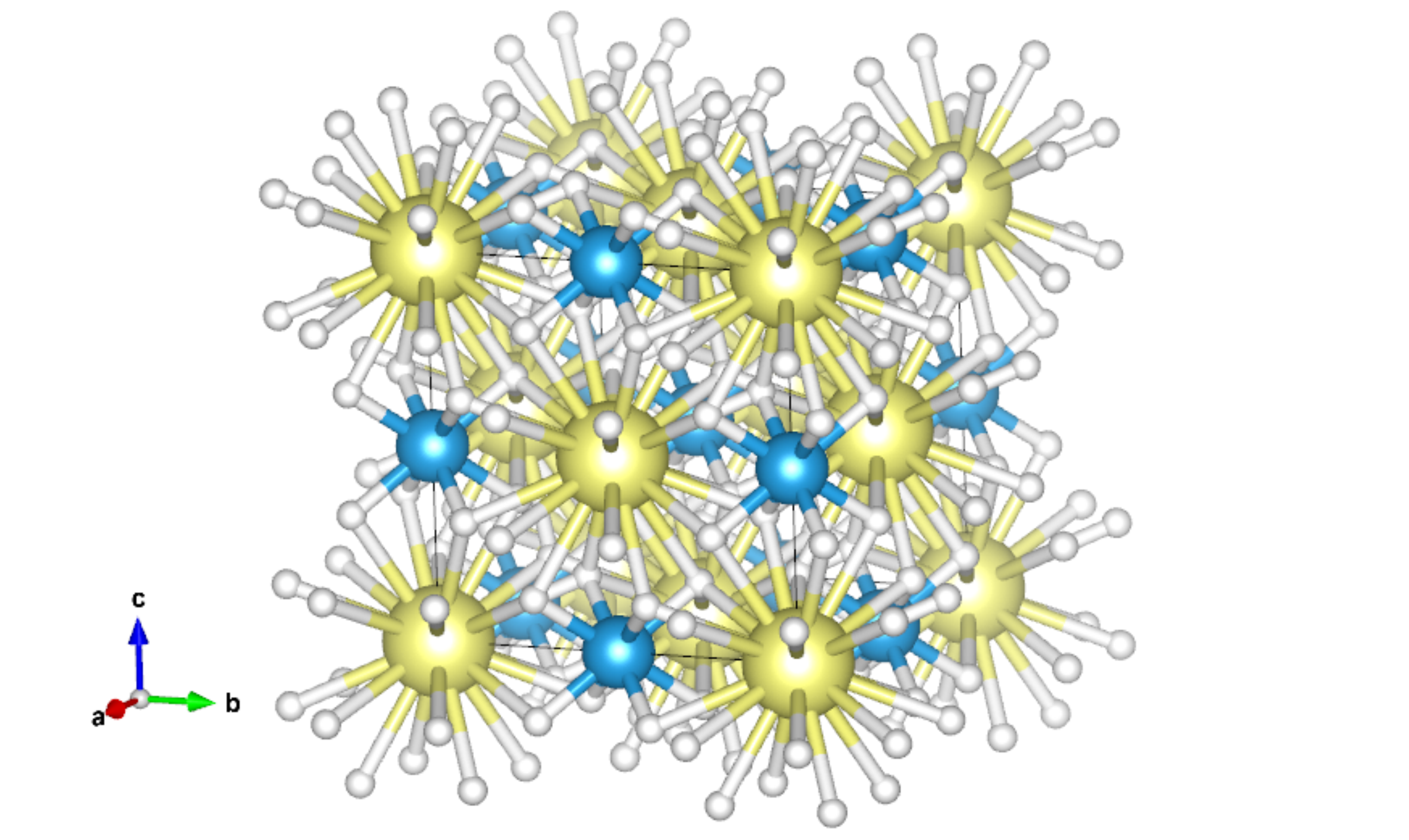}
\caption{The crystal structure of $Fm\overline{3}m$-LuBeH$_8$. The yellow, blue, and white balls represent the lutetium, beryllium, and hydrogen atoms, respectively.}
\label{struct}
\end{center}
   \end{figure*}

The crystal structure of LuBeH$_8$ is shown in Figure 1. The yellow, blue and white balls represent lutetium, beryllium and hydrogen atoms, respectively, with Lu, Be and hydrogen occupying positions 4$a$, 4$b$ , and 32$f$ Wyckoff positions. The H atoms form a polyhedron surrounding the Lu atom. Be atoms are inserted between the polyhedra. $Fm\overline{3}m$-LuBeH$_8$ is structurally similar to sodium hydrides, such as LaH$_{10}$, where guest atoms such as La act as scaffolds and can apply mechanical pressure to the lattice\cite{peng2017hydrogen,wang2012superconductive}, a mechanism commonly referred to as chemical precompression. By linking this mechanism to LuBeH$_8$, the Be atoms occupy the sites between the next closest Lu atoms, effectively filling the remaining interspace in the whole structure. The denser Lu-Be scaffold is then formed, which firmly binds the highly symmetrical metal hydrogen lattice, allowing it to remain stable at lower pressures.

\begin{figure*}
\begin{center}
\includegraphics[width=16cm]{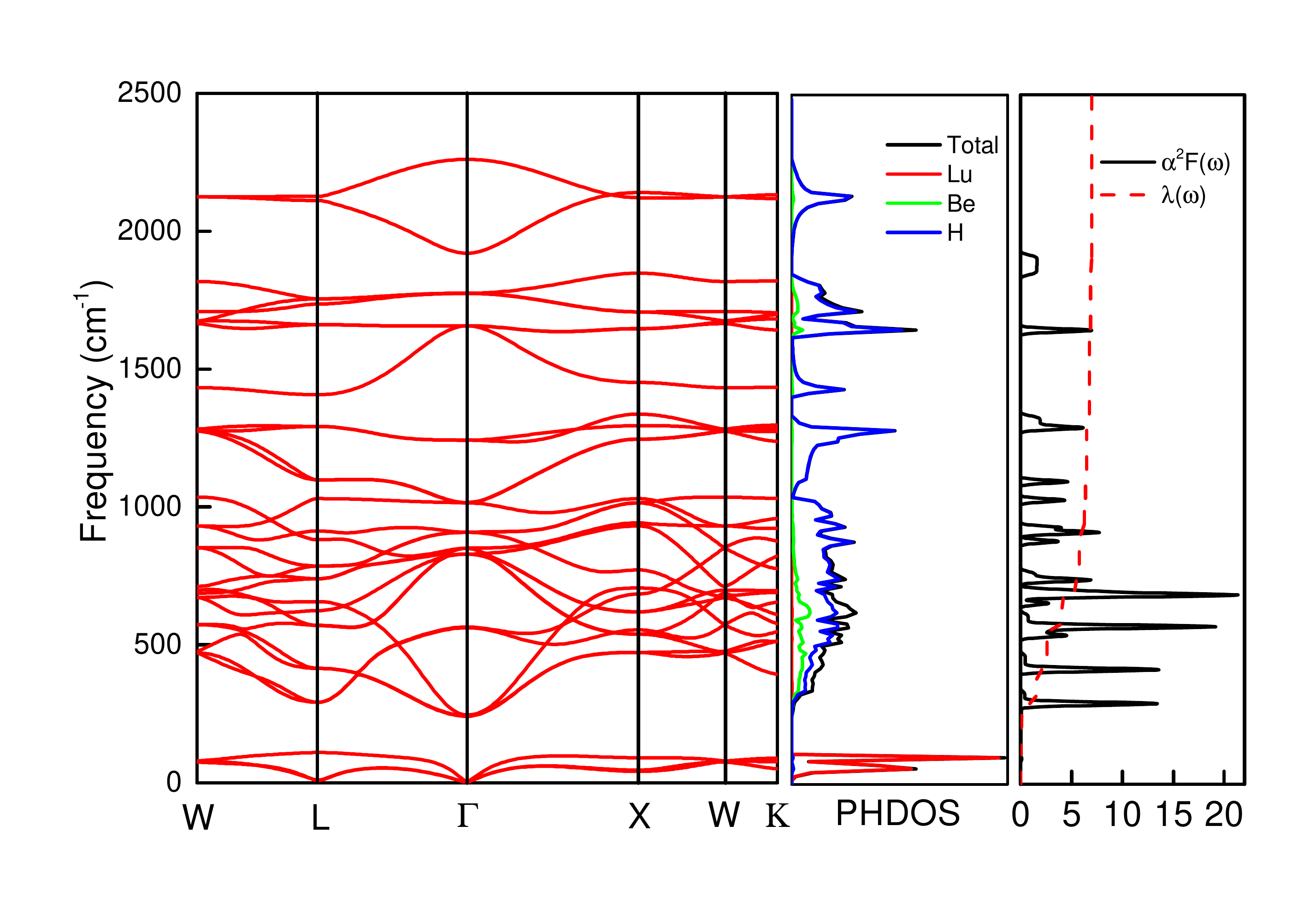}
\caption{$Fm\overline{3}m$-LuBeH$_8$ phonon dispersion, phonon density of states (PHDOS), Eliashberg spectral function $\alpha^2F(\omega)$ and \emph{el-ph} coupling $\lambda(\omega)$ at 100 GPa. The PHDOS projections on Lu, Be, and H are color-coded in red, green, and blue, respectively, to aid in visual interpretation.}
\label{struct}
\end{center}
    \end{figure*}

The phonon properties of $Fm\overline{3}m$-LuBeH$_8$ were calculated using DFPT scheme. The calculation determined that the lower pressure limit of LuBeH$_8$ is 100 GPa, above which no imaginary branches of the phonon spectrum exist. We show the phonon dispersion curve and phonon state density (PHDOS) of LuBeH$_8$ at 100 GPa in Figure 2. It can be seen from the figure that there is no imaginary vibration in the entire Brillouin zone, indicating that the structure is dynamically stable at this pressure, and the vibration of phonons is mainly distributed in the middle and low frequencies in the entire frequency range. From the PHDOS diagram, it can be seen that the vibration in the low frequency region is mainly from the Lu atom, and there is a significant phonon peak located at 100 cm$^{-1}$, and the Be and H atoms in this range are barely vibrating. The vibration in the middle and high frequency range (100 cm$^{-1}$ and above) mainly comes from the H and Be atoms. We also show the integration of the Eliashberg function $\alpha^{2}F(\omega)$ and electron-phonon coupling $\lambda({\omega})$ in the panel on the far right. By integrating the Eliashberg function $\alpha^{2}F(\omega)$, we can get $\lambda=2\int \alpha^2F(\omega)\omega^{-1}d\omega$ and logarithmic mean phonon frequency $\omega_{ln}=exp[2\lambda^{-1}\int d\omega\alpha^{2}F(\omega)\omega^{-1}log\omega]$. According to the coupling curve, it is not difficult to find that the coupling integral below 1200 cm$^{-1}$ accounts for most of the total coupling contribution.

\begin{figure*}
\begin{center}
\includegraphics[width=16cm]{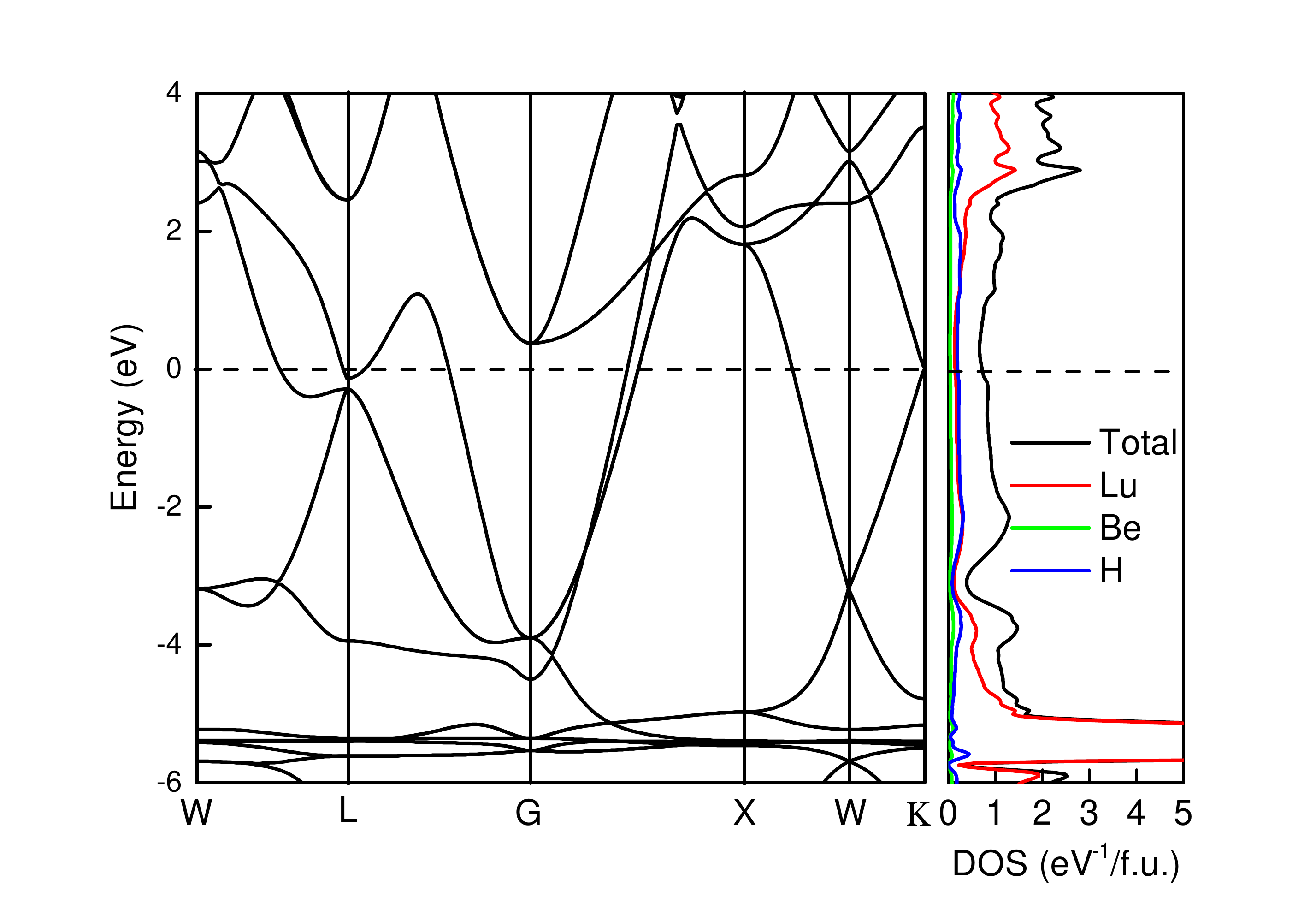}
\caption{$Fm\overline{3}m$-LuBeH$_8$ electronic band structure and partial density of states (DOS) at 100 GPa. The unit of DOS is eV$^{-1}$/f.u., and is color-coded by element, with Lu, Be, and H represented in red, green, and blue, respectively.  The Fermi level serves as the zero point of the energy scale.}
\label{struct}
\end{center}
    \end{figure*}

In Figure 3, we show the electronic band structure of $Fm\overline{3}m$-LuBeH$_8$ at 100 GPa and the atomic projection density of state (DOS) in eV$^{-1}$/f.u. It shows that the structure exhibits metallic behavior, as evidenced by more than one band crossing the Fermi level. Near the Fermi level, DOS is dominated by Lu and H atoms. From the electronic energy band, it can be seen that there is a Dirac-cone like band crossing at point $W$ at $\sim$ -3.2 eV. The corresponding DOS curve around -6 eV has a very high peak, and the Lu atom provides a very large density states due to the $f$ orbital contribution.
\begin{figure*}
\begin{center}
\includegraphics[width=16cm]{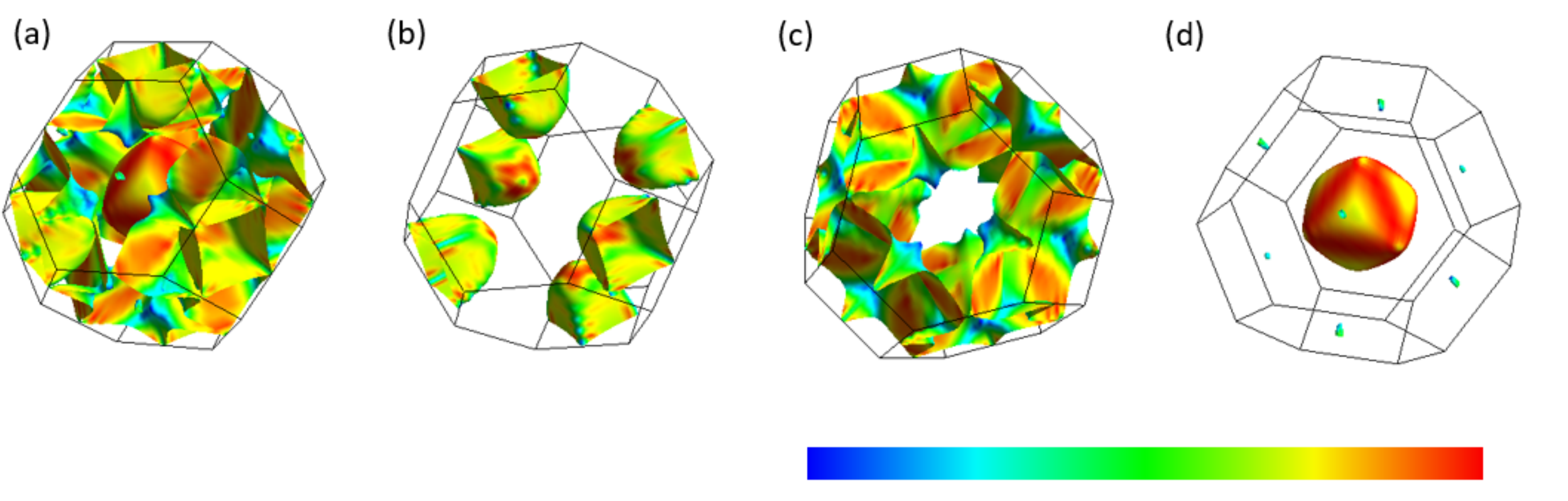}
\caption{The Fermi surfaces of $Fm\overline{3}m$-LuBeH$_8$ at 100 GPa, highlighting the Fermi velocity with a color gradient from blue to red to indicate relative velocity. The overall Fermi surface is shown in (a), and the three distinct parts of the surface are shown in (b)-(d). Each image is color-coded to indicate changes in Fermi velocity.}
\label{struct}
\end{center}
    \end{figure*}

 Figure 4 shows the Fermi surfaces of $Fm\overline{3}m$-LuBeH$_8$, shadowed by the distributions of Fermi velocity, which change color from blue to red to indicate an increase in Fermi velocity. At 100 GPa, the Fermi surface of $Fm\overline{3}m$-LuBeH$_8$ consists mainly of three parts (Figure 4(b-d)), six semi-oval hollow pockets are regularly distributed in the Brillouin area, each pocket is wrapped by a four-leaf clover-shaped sheet with four sharp corners, a large electron sphere around the $\Gamma$ point, and eight small dots regularly around this electron sphere.

  In order to estimate the superconducting critical temperature of LuBeH$_8$ under pressure and to find the maximum $T_{c}$, we performed a linear response calculation for the \emph{e-ph} coupling (EPC), based on the calculated Eliashberg spectral function $\alpha^{2}F(\omega)$,

\begin{equation}
{\alpha ^2}{\rm{F(}}\omega {\rm{) = }}\frac{1}{{2\pi N(0)}}\sum\limits_{Qv}^{} {\frac{{\gamma Qv}}{{\omega Qv}}\delta (\omega  - \omega Qv)},
\end{equation}

the EPC constant $\lambda$ is obtained by:
\begin{equation}
\lambda {\rm{ = }}2\int_0^\infty  {\frac{{{\alpha ^2}{\rm{F}}(\omega )}}{\omega }} d\omega.
\end{equation}

In addition, the critical temperature is calculated by the Allen-Dynes modified Mc-Millan formula\cite{allen1975transition}
\begin{equation}
{T_c} = f1f2\frac{{{\omega _{log}}}}{{1.2}}{\rm{exp}}\left[ { - \frac{{1.04(1 + \lambda )}}{{\lambda  - {\mu ^*}(1 + 0.62\lambda )}}} \right].
\end{equation}

\noindent where $\lambda$ is the EPC intensity, $\omega_{log}$ is the logarithmic mean phonon frequency, and the coulomb pseudopotential parameter $\mu^*$ is set to 0.1.  $\omega_{log}$ is defined as

\begin{equation}
{\omega _{\log }} = \exp [\frac{2}{\lambda }\int_0^\infty  {\frac{{d\omega }}{\omega }{\alpha ^2}{\rm{F}}(\omega )\ln \omega } ].
\end{equation}

The factor $f_1$, $f_2$ depend on $\lambda$, $\mu^*$,$\omega_{log}$, and mean square frequency $\overline{\omega_{2}}$,
\begin{equation}
{f_1}{f_2} = \sqrt[3]{{1 + {{(\frac{\lambda }{{2.46(1 + 3.8{\mu ^*})}})}^{\frac{3}{2}}}}}(1 - \frac{{{\lambda ^2}(1 - \overline{\omega _2}/{\omega _{\log }})}}{{{\lambda ^2} + 3.312{{(1 + 6.3{\mu ^*})}^2}}}).
\end{equation}
The detailed calculation results are shown in Table 1.

\begin{table}[!htbp]
    \centering
    \begin{tabular}{c c c c c}
        \hline
        \hline
       Structure&Pressure (GPa)&$T_{c}$ (K)&$\lambda$&$\omega_{log}$ (K)\\
        \hline
        \multirow{3}*{$Fm\overline{3}m$-LuBeH$_8$}&100&355&7.0&785\\
        &120&293&4.2&900\\
        &150&269&3.0&1100\\
        &180&274.8&2.5&1160\\
        &200&255&2.4&1265\\

        \hline
        \hline
    \end{tabular}
    \caption{The main superconductivity performance of $Fm\overline{3}m$-LuBeH$_8$ at different pressures from 100 to 200 GPa.}\label{Tc}
\end{table}

The calculation results show that $Fm\overline{3}m$-LuBeH$_8$ exhibits metallic behavior while maintaining kinetic stability at 100 GPa, and its superconducting critical temperature is as high as 355 K, which is far beyond the temperature required for room temperature. As the pressure increases, the overall superconducting critical temperature tends to decrease, which may be attributed to the hardening of phonon branches that impede electron-phonon coupling and superconductivity. 

\section{Conclusion}

In conclusion, we discovered the superconducting phase $Fm\overline{3}m$-LuBeH$_8$ of the novel superhydride through first-principles calculations and crystal structure prediction. LuBeH$_8$ remains dynamically stable above 100 GPa while reaching a maximum superconducting critical temperature of 355 K. Excellent superconducting properties result from high structural symmetry and the efficient stacking of beryllium in the lattice, which allows stable mechanical pressure to be applied. $T_{c}$ shows a decreasing tendency with increasing pressure. The increasing pressure will harden the phonon branches and inhibit the electron-phonon coupling of the structure as well as superconductivity. Our study confirms the validity and accuracy of machine-learning-based crystal structure searches and offers a promising approach for discovering other hydride superconductors. Our findings represent a step towards achieving room-temperature superconductivity.

\ack{
This work is supported by the National Key R\&D Program of China (Grant No. 2018YFA0704300), the National Natural Science Foundation of China (Grants No. U1932217), and NUPTSF (Grant No. NY219087, NY220038). Some of the calculations were performed on the supercomputer in the Big Data Computing Center (BDCC) of Southeast University. We thank Prof. Haihu Wen for valuable discussions.\\}

\bibliographystyle{unsrt}
\bibliography{bibfile}

\begin{thebibliography}{10}

\bibitem{van2010discovery}
Dirk Van~Delft and Peter Kes.
\newblock {The discovery of superconductivity}.
\newblock {\em Physics today}, 63(9):38--43, 2010.

\bibitem{bardeen1957theory}
John Bardeen, Leon~N Cooper, and John~Robert Schrieffer.
\newblock {Theory of superconductivity}.
\newblock {\em Physical review}, 108(5):1175, 1957.

\bibitem{ashcroft1968metallic}
Neil~W Ashcroft.
\newblock {Metallic hydrogen: A high-temperature superconductor?}
\newblock {\em Physical Review Letters}, 21(26):1748, 1968.

\bibitem{eremets2008superconductivity}
MI~Eremets, IA~Trojan, SA~Medvedev, JS~Tse, and Y~Yao.
\newblock {Superconductivity in hydrogen dominant materials: Silane}.
\newblock {\em Science}, 319(5869):1506--1509, 2008.

\bibitem{goncharenko2008pressure}
Igor Goncharenko, MI~Eremets, M~Hanfland, JS~Tse, M~Amboage, Y~Yao, and
  IA~Trojan.
\newblock {Pressure-induced hydrogen-dominant metallic state in aluminum
  hydride}.
\newblock {\em Physical Review Letters}, 100(4):045504, 2008.

\bibitem{drozdov2015conventional}
AP~Drozdov, MI~Eremets, IA~Troyan, Vadim Ksenofontov, and Sergii~I Shylin.
\newblock {Conventional superconductivity at 203 kelvin at high pressures in
  the sulfur hydride system}.
\newblock {\em Nature}, 525(7567):73--76, 2015.

\bibitem{zhang2022superconductivity}
CL~Zhang, X~He, ZW~Li, SJ~Zhang, BS~Min, J~Zhang, K~Lu, JF~Zhao, LC~Shi,
  Y~Peng, et~al.
\newblock {Superconductivity above 80 K in polyhydrides of hafnium}.
\newblock {\em Materials Today Physics}, 27:100826, 2022.

\bibitem{kong2021superconductivity}
Panpan Kong, Vasily~S Minkov, Mikhail~A Kuzovnikov, Alexander~P Drozdov,
  Stanislav~P Besedin, Shirin Mozaffari, Luis Balicas, Fedor~Fedorovich
  Balakirev, Vitali~B Prakapenka, Stella Chariton, et~al.
\newblock {Superconductivity up to 243 K in the yttrium-hydrogen system under
  high pressure}.
\newblock {\em Nature communications}, 12(1):5075, 2021.

\bibitem{geballe2018synthesis}
Zachary~M Geballe, Hanyu Liu, Ajay~K Mishra, Muhtar Ahart, Maddury Somayazulu,
  Yue Meng, Maria Baldini, and Russell~J Hemley.
\newblock {Synthesis and stability of lanthanum superhydrides}.
\newblock {\em Angewandte Chemie}, 130(3):696--700, 2018.

\bibitem{drozdov2019superconductivity}
AP~Drozdov, PP~Kong, VS~Minkov, SP~Besedin, MA~Kuzovnikov, S~Mozaffari,
  L~Balicas, FF~Balakirev, DE~Graf, VB~Prakapenka, et~al.
\newblock {Superconductivity at 250 K in lanthanum hydride under high
  pressures}.
\newblock {\em Nature}, 569(7757):528--531, 2019.

\bibitem{liang2019potential}
Xiaowei Liang, Aitor Bergara, Linyan Wang, Bin Wen, Zhisheng Zhao, Xiang-Feng
  Zhou, Julong He, Guoying Gao, and Yongjun Tian.
\newblock {Potential high-T$_c$ superconductivity in CaYH$_{12}$ under
  pressure}.
\newblock {\em Physical Review B}, 99(10):100505, 2019.

\bibitem{sun2019route}
Ying Sun, Jian Lv, Yu~Xie, Hanyu Liu, and Yanming Ma.
\newblock {Route to a superconducting phase above room temperature in
  electron-doped hydride compounds under high pressure}.
\newblock {\em Physical review letters}, 123(9):097001, 2019.

\bibitem{di2021bh}
Simone Di~Cataldo, Christoph Heil, Wolfgang von~der Linden, and Lilia Boeri.
\newblock {LaBH$_8$: Towards high-T$_c$ low-pressure superconductivity in
  ternary superhydrides}.
\newblock {\em Physical Review B}, 104(2):L020511, 2021.

\bibitem{liang2021prediction}
Xiaowei Liang, Aitor Bergara, Xudong Wei, Xiaoxu Song, Linyan Wang, Rongxin
  Sun, Hanyu Liu, Russell~J Hemley, Lin Wang, Guoying Gao, et~al.
\newblock {Prediction of high-T$_c$ superconductivity in ternary lanthanum
  borohydrides}.
\newblock {\em Physical Review B}, 104(13):134501, 2021.

\bibitem{semenok2020distribution}
Dmitrii~V Semenok, Ivan~A Kruglov, Igor~A Savkin, Alexander~G Kvashnin, and
  Artem~R Oganov.
\newblock {On distribution of superconductivity in metal hydrides}.
\newblock {\em Current Opinion in Solid State and Materials Science},
  24(2):100808, 2020.

\bibitem{peng2017hydrogen}
Feng Peng, Ying Sun, Chris~J Pickard, Richard~J Needs, Qiang Wu, and Yanming
  Ma.
\newblock {Hydrogen clathrate structures in rare earth hydrides at high
  pressures: possible route to room-temperature superconductivity}.
\newblock {\em Physical review letters}, 119(10):107001, 2017.

\bibitem{dasenbrock2023evidence}
Nathan Dasenbrock-Gammon, Elliot Snider, Raymond McBride, Hiranya Pasan, Dylan
  Durkee, Nugzari Khalvashi-Sutter, Sasanka Munasinghe, Sachith~E Dissanayake,
  Keith~V Lawler, Ashkan Salamat, et~al.
\newblock {Evidence of near-ambient superconductivity in a N-doped lutetium
  hydride}.
\newblock {\em Nature}, 615(7951):244--250, 2023.

\bibitem{Ming2023}
Xue Ming, Ying-Jie Zhang, Xiyu Zhu, Qing Li, Chengping He, Yuecong Liu,
  Tianheng Huang, Gan Liu, Bo~Zheng, Huan Yang, Jian Sun, Xiaoxiang Xi, and
  Hai-Hu Wen.
\newblock {Absence of near-ambient superconductivity in LuH$_{2\pm{x}}$N$_y$}.
\newblock {\em Nature}, May 2023.

\bibitem{xing2023observation}
Xiangzhuo Xing, Chao Wang, Linchao Yu, Jie Xu, Chutong Zhang, Mengge Zhang,
  Song Huang, Xiaoran Zhang, Bingchao Yang, Xin Chen, Yongsheng Zhang, Jian
  gang Guo, Zhixiang Shi, Yanming Ma, Changfeng Chen, and Xiaobing Liu.
\newblock {Observation of non-superconducting phase changes in
  LuH$_{2\pm{x}}$N$_y$}.
\newblock 2023.

\bibitem{sun2023effect}
Yang Sun, Feng Zhang, Shunqing Wu, Vladimir Antropov, and Kai-Ming Ho.
\newblock {Effect of nitrogen doping and pressure on the stability of cubic
  LuH$_3$}.
\newblock 2023.

\bibitem{hilleke2023structure}
Katerina~P. Hilleke, Xiaoyu Wang, Dongbao Luo, Nisha Geng, Busheng Wang, and
  Eva Zurek.
\newblock {Structure, Stability and Superconductivity of N-doped Lutetium
  Hydrides at kbar Pressures}.
\newblock 2023.

\bibitem{ferreira2023search}
Pedro~P. Ferreira, Lewis~J. Conway, Alessio Cucciari, Simone~Di Cataldo,
  Federico Giannessi, Eva Kogler, Luiz T.~F. Eleno, Chris~J. Pickard, Christoph
  Heil, and Lilia Boeri.
\newblock {Search for ambient superconductivity in the Lu-N-H system}.
\newblock 2023.

\bibitem{wu2022phase}
Sixuan Wu, Bin Li, Zhi Chen, Yu~Hou, Yan Bai, Xiaofeng Hao, Yeqian Yang,
  Shengli Liu, Jie Cheng, and Zhixiang Shi.
\newblock {Phase transitions and superconductivity in ternary hydride Li2SiH6
  at high pressures}.
\newblock {\em Journal of Applied Physics}, 131(6):065901, 2022.

\bibitem{crystree2}
Bin Li.
\newblock {Crystree: a crystal structure predictor based on machine learning,
  https://github.com/bliseu/CRYSTREE}.
\newblock 2020.

\bibitem{nwad128}
{Wang, Junjie and Gao, Hao and Han, Yu and Ding, Chi and Pan, Shuning and Wang,
  Yong and Jia, Qiuhan and Wang, Hui-Tian and Xing, Dingyu and Sun, Jian}.
\newblock {MAGUS: machine learning and graph theory assisted universal
  structure searcher}.
\newblock {\em National Science Review}, 05 2023.
\newblock nwad128.

\bibitem{giannozzi2009quantum}
Paolo Giannozzi, Stefano Baroni, Nicola Bonini, Matteo Calandra, Roberto Car,
  Carlo Cavazzoni, Davide Ceresoli, Guido~L Chiarotti, Matteo Cococcioni,
  Ismaila Dabo, et~al.
\newblock {QUANTUM ESPRESSO: a modular and open-source software project for
  quantum simulations of materials}.
\newblock {\em Journal of physics: Condensed matter}, 21(39):395502, 2009.

\bibitem{baroni2001phonons}
Stefano Baroni, Stefano De~Gironcoli, Andrea Dal~Corso, and Paolo Giannozzi.
\newblock {Phonons and related crystal properties from density-functional
  perturbation theory}.
\newblock {\em Reviews of modern Physics}, 73(2):515, 2001.

\bibitem{blaha1990full}
Peter Blaha, Karlheinz Schwarz, P~Sorantin, and SB~Trickey.
\newblock {Full-potential, linearized augmented plane wave programs for
  crystalline systems}.
\newblock {\em Computer physics communications}, 59(2):399--415, 1990.

\bibitem{momma2011vesta}
Koichi Momma and Fujio Izumi.
\newblock {VESTA 3 for three-dimensional visualization of crystal, volumetric
  and morphology data}.
\newblock {\em Journal of applied crystallography}, 44(6):1272--1276, 2011.

\bibitem{kawamura2019fermisurfer}
Mitsuaki Kawamura.
\newblock {FermiSurfer: Fermi-surface viewer providing multiple representation
  schemes}.
\newblock {\em Computer Physics Communications}, 239:197--203, 2019.

\bibitem{wang2012superconductive}
Hui Wang, John~S Tse, Kaori Tanaka, Toshiaki Iitaka, and Yanming Ma.
\newblock {Superconductive sodalite-like clathrate calcium hydride at high
  pressures}.
\newblock {\em Proceedings of the National Academy of Sciences},
  109(17):6463--6466, 2012.

\bibitem{allen1975transition}
Ph~B Allen and RC~Dynes.
\newblock {Transition temperature of strong-coupled superconductors
  reanalyzed}.
\newblock {\em Physical Review B}, 12(3):905, 1975.

\end{thebibliography}

\end{document}